\documentclass[sigconf]{acmart}
\pdfoutput=1
\usepackage{xspace}
\usepackage{enumitem}

\newcommand{\kmer}{$k$-mer\xspace} 
\newcommand{\ours}{HySortK\xspace} 
\newcommand{\kmers}{$k$-mers\xspace} 
\newcommand{\mmer}{$m$-mer\xspace} 
\newcommand{\mmers}{$m$-mers\xspace} 
\newcommand{\supermer}{$supermer$\xspace}
\newcommand{\supermers}{$supermers$\xspace}
\newcommand{\minimizer}{$minimizer$\xspace}

\newcommand{\kmerlist}{$kmerlist$\xspace}
\newcommand{\deque}{\texttt{deque}\xspace}

\AtBeginDocument{%
  }

\setcopyright{acmlicensed}
\copyrightyear{2024}
\acmYear{2024}
\setcopyright{rightsretained}
\acmConference[ICPP '24]{The 53rd International Conference on Parallel Processing}{August 12--15, 2024}{Gotland, Sweden}
\acmBooktitle{The 53rd International Conference on Parallel Processing (ICPP '24), August 12--15, 2024, Gotland, Sweden}
\acmPrice{}
\acmDOI{10.1145/3673038.3673072}
\acmISBN{979-8-4007-1793-2/24/08}

\begin{document}

\title{High-Performance Sorting-Based \kmer Counting in Distributed Memory with Flexible Hybrid Parallelism}

\author{Yifan Li}
\email{yf-li21@mails.tsinghua.edu.cn}
\affiliation{%
  \institution{Tsinghua University}
  \streetaddress{30 Shuangqing Rd}
  \city{Beijing}
  \country{China}
  \postcode{43017-6221}
}

\author{Giulia Guidi}
\email{gguidi@cornell.edu}
\affiliation{%
  \institution{Cornell University}
  \streetaddress{107 Hoy Rd}
  \city{Ithaca}
  \state{NY}
  \country{USA}}

\renewcommand{\shortauthors}{Li and Guidi}

\begin{abstract}

In generating large quantities of DNA data, high-throughput sequencing technologies require advanced bioinformatics infrastructures for efficient data analysis. 
\kmer counting, the process of quantifying the frequency of fixed-length $k$ DNA subsequences, is a fundamental step in various bioinformatics pipelines, including genome assembly and protein prediction.
Due to the growing volume of data, the scaling of the counting process is critical.

In the literature, distributed memory software uses hash tables, which exhibit poor cache friendliness and consume excessive memory.
They often also lack support for flexible parallelism, which makes integration into existing bioinformatics pipelines difficult.
In this work, we propose \ours, a highly efficient sorting-based distributed memory \kmer counter.
\ours reduces the communication volume through a carefully designed communication scheme and domain-specific optimization strategies.
Furthermore, we introduce an abstract task layer for flexible hybrid parallelism to address load imbalances in different scenarios.

\ours achieves a 2-10$\times$ speedup compared to the GPU baseline on 4 and 8 nodes.
Compared to state-of-the-art CPU software, \ours achieves up to 2$\times$ speedup while reducing peak memory usage by 30\% on 16 nodes.
Finally, we integrated \ours into an existing genome assembly pipeline and achieved up to $1.8\times$ speedup, proving its flexibility and practicality in real-world scenarios.

\end{abstract}





\maketitle

\section{Introduction}

A \kmer, i.e. a substring of fixed length $k$, plays a crucial role in various bioinformatics pipelines~\cite{hofmeyr2020terabase, guidi2021parallel, selvitopi2020distributed, nurk2020hicanu}.
$K$-mer counting is a common operation where we count the frequency of \kmers in a given dataset of DNA sequences, also known as \emph{reads}.
Usually, the distribution of \kmers or \kmers with frequencies in a certain range is collected for further analysis.
In genome assembly and pangenomics analysis, for example, \kmers are used to identify potential seed matches between sequences for subsequent fine-grained alignment, commonly known as the ``seed and extend'' pattern~\cite{yan2021accel, guidi2021parallel, rautiainen2020graphaligner}.
Furthermore, statistical studies, including recent machine learning approaches~\cite{doi:10.1021/acs.jpclett.3c02817}, treat \kmers as features in datasets for learning and inference purposes. 
Precise and efficient counting of \kmer frequencies is at the core of all these applications.

In recent years, advances in high-throughput sequencing technologies have led to an exponential increase in the size of genomic datasets.
In particular, long-read sequencing technologies, characterized by a longer sequence (read) length, have gained popularity.
This increase in data promotes new applications or studies with greater precision but also poses a challenge for bioinformatics software.
The size of input data frequently surpasses the memory capacity of a single machine, resulting in performance degradation when relying on I/O performance or, in some cases, out-of-memory failures.
The \kmer counting task is sensitive to the growing amount of data, as this is usually the first stage of a workflow and filtering the input data is often not an option~\cite{georganas2015hipmer,baaijens2022computational,breitwieser2018krakenuniq}.
Therefore, there is an urgent need for efficient \kmer counters with distributed memory.

However, parallelizing \kmer counting in distributed memory is a non-trivial task.
The runtimes of MetaHipMer2, a short-read \emph{de novo} metagenome assembler, show that the \kmer counting stage can take almost 50\% of the total time~\cite{awan2021accelerating}.
In contrast to other stages, it is not possible to partition the input data directly into non-contiguous regions of the genome and treat the \kmer count in an embarrassingly parallel manner.
Due to the short \kmer length, the error rate of the experimental sequencing process, and the intrinsic repetitiveness of biological sequences, there is no guarantee that the instances of a \kmer subsequence will appear in a precise and individual part of the generated dataset~\cite{guidi2021bella}.
Both data structure and algorithms must be adapted for a distributed memory scenario.
Crucially, when leveraging the computing power of multiple machines or nodes, a substantial portion of the total time is spent on communication.
Optimizing communication is critical.

To overcome these challenges, we propose a novel radix sort-based distributed memory \kmer counter: \ours\footnote{Our code is available at\url{https://github.com/CornellHPC/HySortK}}.
To date, most distributed memory software has relied on hash tables to determine the\kmer frequency~\cite {hofmeyr2020terabase, guidi2021parallel, nisa2021distributed, pan2018optimizing}.
Yet, concurrent access to hash tables can be difficult, they consume more memory and are not very efficient due to their random memory access pattern.
In contrast, we take a different approach by using array data structures for \kmer storing and relying on radix sort and linear scan to determine their frequency.
Radix sorting can be performed in-place, reducing peak memory usage and eliminating the need for the Bloom filter.
This eliminates the need for a round of data exchange and significantly reduces communication time.
In addition, we introduce a domain-specific compression method to reduce the communication volume.
To maintain high performance and support flexible hybrid parallelization (i.e., any number of MPI processes and OpenMP threads per processor), we introduce an abstraction layer of tasks between MPI processes and OpenMP threads.
Furthermore, with task-based parallelization, we can effectively address load-balancing issues.
Our main contributions include:
\vspace{-.05em}
\begin{itemize}

\item{Here, we redesign the distributed \kmer counting task as a sort-based problem instead of a hash table-based problem, which reduces random memory access and memory usage.
In \ours, we improve the supermer strategy and use a hash function to ensure order.
Our supermer partitioning introduces a new method to find minimizers for consecutive \kmers in a genomic sequence.
}

\item{\ours uses flexible hybrid MPI and OpenMP parallelization so that it can be seamlessly integrated into existing pipelines with improved performance.
\ours was integrated into a \emph{de novo} long-read genome assembly workflow.
The hybrid approach enables us to introduce a task abstraction layer to address load imbalances in specific scenarios.
}

\item {\ours is $2$-$10\times$ faster than GPU approaches.
It has shown up to $2\times$ speedup and up to $3\times$ lower memory usage compared to state-of-the-art CPU software.
Finally, integration into a distributed memory genome assembly workflow has resulted in an end-to-end speedup of up to $1.8\times$.}

\end{itemize}

\vspace{-.5em}
\section{Background}\label{sec:background}

In this section, we describe \kmer counting, common parallel approaches to its implementation, and existing parallel sorting approaches.
Finally, we describe the supermer optimization strategy.

\vspace{-.5em}
\subsection{$k$-mer Counting}
%
The \kmer counting process involves counting the occurrence of subsequences with a length of $k$ nucleotides (A, C, T, G) in genomic or transcriptomic data.
It is common for a genomics pipeline to begin by enumerating \kmer frequencies in the input data~\cite{nurk2020hicanu, jonkheer2022pantools}.
The resulting \kmer histogram is critical for understanding the distribution of genomic subsequences, profiling genomic and metagenomic data, identifying scientifically relevant \kmers based on their abundance, and more.
The histogram serves as the basis for various downstream analyses, such as the representation of a de Bruijn graph~\cite{georganas2015hipmer}, the generation of large-scale sequence search indices~\cite{jonkheer2022pantools}, and weighted locus locality-sensitive hashing~\cite{marccais2019locality}.

\begin{figure*}[h]  
  \centering  
  \includegraphics[width=.885\textwidth]{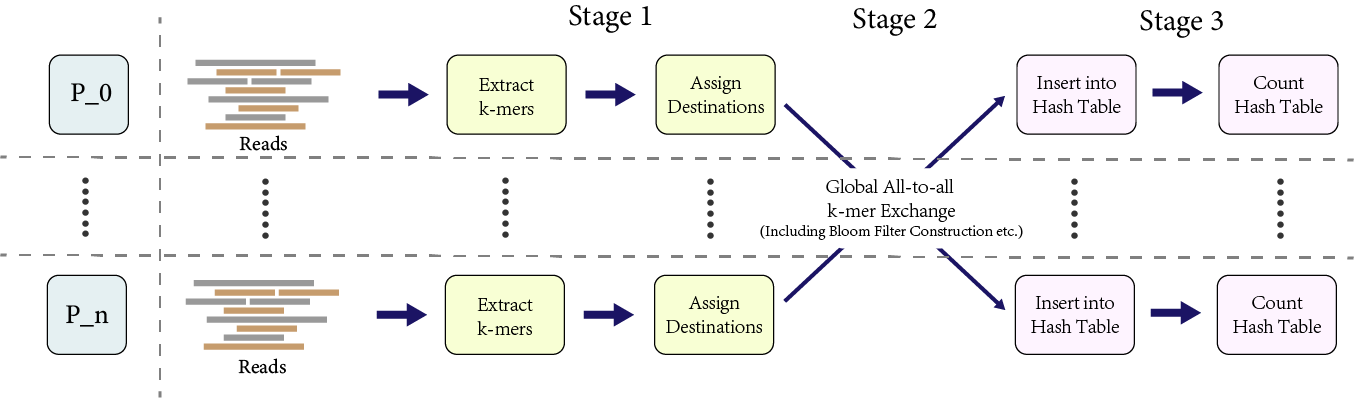}  
  \vspace{-.5em}
  \caption{Overview of the common paradigm for distributed memory \kmer counting pipelines using hash-tables.}  
  \Description{}
  \label{fig:distributed_kmer_counting}  
  \vspace{-1em}
\end{figure*}


The input irregularity due to biology and the experimental sequencing process poses a major challenge for the parallelization of \kmer counting in distributed memory.
In particular, the distribution of \kmers in the biological input is not fixed and can only be determined at runtime.
In the literature, scalable distributed memory \kmer counting mainly relies on hash tables.

\vspace{-.5em}
\subsection{Parallel \kmer Counting}\label{sec:backgroundp}

This section provides an overview of common practices for counting \kmers in distributed memory.
Common distributed \kmer counter software, including DEDUKT~\cite{nisa2021distributed}, a GPU-based tool, the \kmer counting module of ELBA~\cite{guidi2021parallel}, a distributed memory \emph{de novo} long-read genome assembler, and the \kmer counting module of PaKman~\cite{ghosh2020pakman} follow the general approach proposed by Georganas et al.~\cite{georganas2014parallel}.
It includes three main stages: (1) local reading and parsing of \kmers, (2) exchange of \kmers, and (3) local counting of \kmers.
In addition, DEDUKT uses the \supermer concept to reduce communication.
The process of encoding and decoding \kmers is added to the first and last stages, and the exchange of \kmers is replaced by the exchange of supermers in the second stage.

In Figure~\ref{fig:distributed_kmer_counting}, the sequences from the input file are divided evenly between the processes using a greedy algorithm.
The processes then independently read batches of sequences, encode the sequences, and parse them into \kmers (1).
The processes use the same hash function and mod operation as a mapping function to compute an identifier (ID) for each \kmer and divide the local \kmer set into groups based on such function.
The uniform hash function ensures that \kmers with the same value have the same identifier.

The second stage (2) is the most communication-intensive, in which the \kmers are distributed in groups to different target processes based on their respective IDs.
Depending on the application, we may want to exchange only the \kmer itself or both the \kmer and its \emph{extension} information.
The extension information of a \kmer includes the ID of the read (i.e., a DNA sequence) from which the \kmer was extracted and the position of the \kmer in this read.
This stage, which resembles an all-to-all exchange, can extend over several rounds due to memory constraints or a large number of processes.
A common strategy is a two-pass approach that aims to filter out erroneous \kmers and reduce memory footprint. 
This often comes at the cost of increased communication. 
Before the first communication run, where \kmers are exchanged, a Hyperloglog data structure~\cite{flajolet2007hyperloglog} is used to estimate the number of unique \kmers instances.
Hyperloglog data structures are first built locally and then merged globally.
The communication volume during this phase does not depend on the size of the data set but is a function of the parameter $k$.
The time required is negligible and we do not consider it as a pass, even though it is referred to as the first pass in Georganas et al.~\cite{georganas2014parallel}.
Once the number of unique \kmers is estimated, a Bloom filter is constructed accordingly. 
In the first pass, \kmers are exchanged without the extension information and inserted into the Bloom filter of the target process.
In the second pass, \kmers are then exchanged together with the extension information, if this information is needed.
In the target process, the hash table created in the first pass is used to filter out singletons, which are considered to be the result of sequencing errors.

In the third stage (3), the counting takes place, which requires no additional communication.
Using a hash table-based approach, the \kmers coming through the Bloom filter in the second exchange pass are often inserted into a temporary array and then inserted into the hash table. 
The hash table serves as a counter, recording the value of the \kmer, its appearance frequency, and the possible extension information.
Each local hash table entry corresponds to a <key, value> pair, where the key is the \kmer itself and the value is the frequency.
The \kmer value must be stored as collisions are possible and correctness should be ensured if the chaining or open addressing method is used.
The method described above can generally be regarded as a distributed hash table approach.

\subsection{Parallel Sorting}\label{sec:parallelsort}

The exponential growth of data in various areas has made more efficient sorting methodologies necessary.
As modern hardware is mostly composed of multicore architectures, the literature focuses on the use of thread-level parallelism to increase performance.

In most cases, one of two strategies is used for multicore sorting: top-down or bottom-up \cite{cho2015paradis}.
In the first strategy, the input dataset is first partitioned based on the key, with each group then sorted independently.
In contrast, the second approach divides the dataset without a specific rule to facilitate load balancing and the results are merged after each partition has been sorted independently.

The radix sort algorithm is well suited for multicore parallelization \cite{kokot2017sorting}.
Its theoretical complexity is $\mathcal{O}(n \cdot d)$, which makes it more efficient for large input data than comparison-based algorithms, where $n$ is the number of items to be sorted and $d$ is a constant dependent on $k$.
In contrast, comparison-based algorithms have a higher lower bound of $\mathcal{O}(n\log n)$. PARADIS \cite{cho2015paradis} is a parallel in-place radix sort algorithm known for its low memory footprint and its use of adaptive load balancing to minimize overall processing time. 
RADULS, another parallel radix sort algorithm, was developed for the efficient management of very large datasets \cite{kokot2017sorting}.
It is cache-friendly optimized for modern hardware and requires more physical memory as it is not an in-place algorithm.
RADULS has been successfully implemented in a shared memory \kmer counter~\cite{kokot2017kmc}.
Our \kmer counter, \ours, uses both PARADIS and RADULS and switches between the two depending on the available memory.

\vspace{-.5em}
\subsection{Supermer}

The \kmer counting problem makes it possible to adopt some domain-specific optimization strategies, such as the \supermer concept.
A \supermer, or super \kmer, is a contiguous sequence of DNA bases. 
The length of each supermer is greater than or equal to $k$, the length of \kmers.
It is important to note that \kmers extracted from the supermer should have the same target process.
The overlapping subsequence of these \kmers is not exchanged repeatedly, which reduces the communication volume many times over.
However, the na\"ive way of assigning \kmers to processes can lead to a low probability that adjacent \kmers belong to the same process.
Therefore, \mmer and \minimizer are proposed to address this problem.

A \mmer is defined as a length $m$ subsequence of DNA bases, where $m$ is smaller than $k$.
Each \kmer thus includes multiple \mmers.
A minimizer is the \mmer of a \kmer with the lowest score for a function $f$.
The target process of the \kmer is no longer determined by the hash value of the \kmer itself but by the hash value of the minimizer.
Consecutive \kmers are likely to share the same minimizer, which increases the probability that adjacent \kmers will be assigned to the same target process.
As a result, supermers reduce the data exchange volume many times over.

The supermer concept was originally proposed by MSP~\cite{li2015mspkmercounter}.
Later \kmer counters, such as KMC3~\cite{kokot2017kmc} and DEDUKT~\cite{nisa2021distributed}, have also adopted this approach.
The literature has focused on choosing a reasonable score function $f$.
Computational complexity and load balance are the main concerns when choosing a score function.
Common score functions include modified lexical ordering and randomized arithmetic computation.

\section{Methodologies}\label{sec:method}

In this section, we describe our approach and the optimization strategies we have implemented.
Our approach has four main differences compared to state-of-the-art CPU-based software:

\begin{enumerate}[label=(\alph*)]
    \item {First, the method for counting \kmers is changed from a hash-table-based approach to a sorting-based approach;}

    \item{Our approach uses the \supermer concept to reduce the volume of data exchange.
    To improve the traditional supermer approach~\cite{li2015mspkmercounter}, we use a hash function to determine the order of \mmers and a linear method to find the minimizers of consecutive \kmers independent of the length $k$.
    The optimized supermer strategy improves load balance with negligible additional computation.}

    \item {By combining the sort-based approach and careful supermer optimization, \ours significantly reduces the memory footprint, making a BLOOM filter superfluous. 
    Consequently, the exchange no longer takes place in two passes.
    Each \kmer is exchanged at most once, which further reduces the communication volume.
    }
    \item {
    \ours uses a task abstraction layer to enable flexible hybrid parallelization with OpenMP and MPI.
   In addition, the task design enables the detection of frequent \kmers (or \emph{heavy hitters}) with low overhead,
 while providing efficient methods to address load imbalance.}
\end{enumerate}

\subsection{Sorting-Based \kmer Counting}

A key component of \ours is the sorting-based approach, which is both flexible and efficient.
Similar strategies have been identified in KMC3~\cite{kokot2017kmc}, but to the best of our knowledge not for distributed memory.
The latest version of kmerind ~\cite{pan2018optimizing} introduces a sorting-based approach.
However, it is based on inter-process sample sort and is slower than kmerind based on hash tables.

In this section, we present our approach, address the challenges arising from the popular hash table approach, and explain how our approach addresses these challenges.

The backbone of \ours is similar to the approach described in Section~\ref{sec:backgroundp}. 
In the first stage, the sequences are parsed locally into \kmers, in the second stage \kmers are exchanged across processes, and in the third stage, the \kmers are counted locally in each process.
The differences between the hash table approach and the sorting approach are mainly consolidated in the third stage.

In particular, the \kmers remain in the receive buffer after the exchange phase and wait for the local count.
\ours does not create redundant copies as it computes a parallel multithreaded in-place radix sort to reorder the \kmer instances according to the value directly in the buffered data.
Only when more memory resources are available, \ours switches to a more efficient radix sort algorithm that requires an auxiliary array for counting.
In both cases, it is guaranteed that \kmers with the same value fall into adjacent places after sorting.
The threads scan the sorted array to count the time of occurrence of each \kmer, which is a linear time complexity operation.
Depending on the application, we can record only the final counting histogram or copy the \kmers that meet our requirements to another array for further indexing and querying.

Our approach relies on the efficiency of the sort algorithm.
Radix sort is chosen because its time complexity is $\mathcal{O}(n\cdot d)$, where $n$ is the size of the \kmer array and $d$ is a constant with respect to $k$, the length of each \kmer.
In \ours, we use the PARADIS and RADULS sorting algorithm, as proposed by ~\cite{cho2015paradis} ~\cite{kokot2017sorting}.
RADULS is faster but uses more memory.
In \ours, the processes read the system status after the exchange phase to estimate the available memory.
If sufficient memory is available for out-of-place sorting, RADULS is used, otherwise PARADIS.


Hash tables often suffer from slow performance and high memory usage, as they require up to twice the memory of the stored data due to the additional structures required and a standard load factor of 0.7~\cite{10.14778/2850583.2850585}.
Efforts to reduce memory usage include filtering out infrequent elements (e.g., singletons) and using Counting Bloom Filter~\cite{ge2020counting}, however, these strategies may limit functionality or accuracy, respectively, and may not be suitable for every application.
Conversely, our sorting approach hardly requires any additional space when memory resources are limited.
\ours eliminates the need to discard singletons, thereby avoiding the earlier steps of constructing the Bloom filter, saving considerable time while still achieving superior memory efficiency.

Concurrency is a major challenge for hash tables when numerous insert operations are required, such as when \kmer counting.
One solution is to manage the shared memory in a kind of distributed memory, where each thread is responsible for a partition of \kmers.
Pan et al.~\cite{pan2018optimizing} chose this approach to reduce communication time, as it theoretically does not involve overhead.
Empirically, however, this approach has proven to be inefficient, as both Pan et al. and this work in Section~\ref{sec:results} show.
The sorting approach facilitates multithreading scaling.
Using less than or equal to 16 threads in our evaluation, linear or near-linear scaling can be achieved for parallel sorting algorithms.
The task abstraction layer we introduced provides further relief.
The overall performance gain from our optimization strategies is $2-5\times$ compared to a baseline implementation with hash tables.

\vspace{-.5em}
\subsection{Optimized Supermer Strategy}

Communication between nodes is often a bottleneck in distributed memory \kmer counting.
The supermer strategy presented in Section~\ref{sec:background} is therefore used to reduce the communication volume.
In \ours, we adopt and improve this strategy by introducing a more efficient method for finding minimizers of consecutive \kmers.

\begin{figure}[t]  
  \centering  
  \includegraphics[width=\columnwidth]{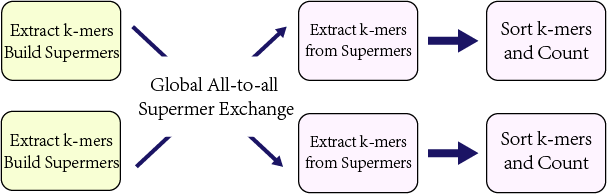}  
  \caption{Overview of sorting-based \kmer counting using the supermer strategy.}  
  \Description{}
  \label{fig:sort_plus_supermer}  
\end{figure}

In \ours, we use the same hash function to determine the minimizers of \kmers and the corresponding destination process for each \minimizer.
The hash value of each \mmer serves as its score, where the \mmer with the lowest value is called the minimizer.
The remainder of the hash value, divided by the number of processes, determines the target for \kmers associated with that minimizer.
The goal is to balance the number of \kmers going to each target to avoid load imbalance and longer execution times.
Both the selection of the minimizer and the decision on the target based on the minimizer are critical to the performance of the supermer strategy.

The randomness of the hash function makes it a suitable choice for the score function.
A variety of hash functions are available, and empirically, we find that a single hash function suffices for both finding minimizers and deciding destinations.
Research on the use of universal hash functions~\cite{carter1977universal} as scoring functions is limited because these functions can assign the same hash value to different elements, which can lead to different minimizers and thus target processes being assigned to \kmers with the same value.
In addition, they usually require more computation than simpler methodologies.

To solve the first problem, we use the same hash function to search for minimizers and to decide on the target to ensure correctness even if different minimizers are assigned to \kmers with the same value.
In the rare case that multiple \mmers of a \kmer collide in their hash value, the \kmer will be assigned to the same target regardless of which \mmer is chosen as the minimizer.
Even if two different hash functions are used, a simple strategy can be applied to eliminate all potential errors.
If multiple \mmers of a single \kmer have the same hash value, a secondary comparison using the lexical order can eliminate the uncertainty in the choice of minimizer and effectively solve the problem.

To address the second problem, we choose a hash function that requires a reasonable computation and propose a method to efficiently find minimizers.
\texttt{Murmurhash3}, previously used for hashing \kmers, is used due to its randomness and low computational requirement.
Computing the hash value of an \mmer for minimizer saves computation in the decision process for the target.
Our evaluation shows that using a hash function can lead to an increase in runtime of up to $2\times$ compared to a simple scoring function such as lexical ordering.
However, the overhead is negligible when considering the overall runtime of the pipeline.

Furthermore, we optimize the strategy to find minimizers of consecutive \kmers.
Let us assume the length of a read (i.e., a DNA sequence) is $n$, and we want to identify minimizers for its \kmers. DEDUKT~\cite{nisa2021distributed} computes the \mmers for each \kmer independently,  which can result in redundant computation. 
The sliding window concept introduced by Li et al.~\cite{li2015mspkmercounter} for finding minimizers involves scanning through the \kmers in a read while preserving the score and position of the current minimizer.
The minimizer is updated if a better score is found; otherwise, it's retained.
This method, though efficient, requires recalculating the valid minimizers when the current one has ``expired'', which can result in a high computation.

Here, we implement an improved sliding window maximum value algorithm.
First, the read operation is performed as described by Li et al. \cite{li2015mspkmercounter}, but instead of keeping a single variable for the current \minimizer, a \deque is kept.
Each element in the \deque stores the score and position of a valid \mmer. 
The elements in the \deque are ordered monotonically, i.e. the \mmers in the \deque have increasing scores, which ensures that the front of the \deque stores the current \minimizer.
So when we move the sliding window forward, we add at most one new \mmer to the \deque and remove at most one.
To remove an expiring \mmer in the \deque, we check the front part of the \deque. 
If the front element is the expiring \mmer, it is removed; otherwise nothing is done.
To insert an \mmer, we remove elements at the end of the \deque until the score of the end element is lower than that of the new \mmer or the \deque is empty, and then insert the new \mmer at the end.
Correctness is ensured by rule (a), which removes expired \kmers, and rule (b), which stores \kmers in order within the \deque.
The sliding window only saves valid minimizers, whereby the front element of the \deque is always the current minimizer.
Thus, new entries either have a lower score or replace less relevant entries, which ensures that the minimal element remains at the front.
Removed entries are replaced by better minimizers that appear later.

Our \supermer strategy resulted in $80\%$ less communication and a $3\times$ speedup in the exchange phase at $k=31$.
Using \texttt{Murmurhash3} as a scoring function on a 31 GB \emph{H. sapiens} dataset, we obtained a balanced partitioning of 256 batches with a standard deviation of batch size more than $10\ times$ smaller than the standard partitioning of KMC3, and a max-to-min ratio of only $1.06$.



\begin{figure*}[h]  
  \centering  
  \includegraphics[width=.885\textwidth]{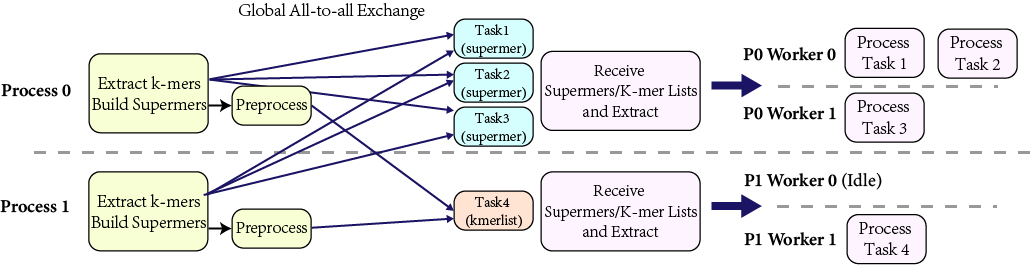}  
  \vspace{-.5em}
  \caption{Overview of \ours paradigm for distributed memory \kmer counting with hybrid task parallelism.}  
  \Description{}
  \label{fig:hysortk}  
\end{figure*}

\subsection{Communication Optimization}
\label{sec:comm_optimize}

\ours reduces communication by replacing the two-pass approach with a one-pass approach and applying the \supermer strategy.
\ours also introduces two strategies to further reduce communication time in the second stage in Figure~\ref{fig:distributed_kmer_counting}, i.e. the most communication-intensive stage. 
The first strategy is to overlap communication with computation while the second uses data compression.
Our data compression does not result in any data loss.
\ours provides exact \kmer count.

\subsubsection{Communication and computation overlap}
Communication is divided into rounds.
Each process is allowed to send a limited number of bytes to other processes in a communication round.
In \ours, we use \texttt{Alltoall} instead of \texttt{Alltoallv} because the regular pattern of \texttt{Alltoall} generally provides better performance.
However, when using \texttt{Alltoall}, padding is required, which can lead to longer communication times if the partition of \supermers is unbalanced.
The non-blocking version \texttt{Ialltoall} is called to support the overlap.
Two send and receive buffers are allocated for each process.
The pointers to these buffers are swapped after each round.
Therefore, the data exchange for round $n$ can take place simultaneously with the parsing of the receive buffer for round $n-1$ and the preparation of the send buffer for round $n+1$.

\subsubsection{Data compression}
In genomics applications, such as genome assembly~\cite{guidi2021parallel}, information about the extension of \kmers is sometimes required for downstream analysis.
Common information includes the \texttt{read\_id} and \texttt{pos\_in\_read} (i.e., location at which the \kmer is found in the sequence) for each \kmer.
Under a reasonable choice of $k$ (typically less than 63), this information consumes more bytes than the \kmer value itself.
Compression of this extension information is therefore highly beneficial.
In \ours we compress this information using domain-specific knowledge.
The length of long read sequences is usually between $1000$ and $20,000$ bases~\cite{logsdon2020long}, much larger than the total number of processes in most scenarios.
Consequently, there is a high probability that consecutive \kmers going to the same destination process will have the same \texttt{read\_id}, and the difference in \texttt{pos\_in\_read} is usually small and in the range of a \texttt{int8}.
So instead of using a \texttt{int32} field to capture the \texttt{pos\_in\_read},
we use a smaller \texttt{int8} field to capture the difference of its position to the last \kmer travelling to the same destination.
The same strategy applies to the \texttt{read\_id} field.
An additional byte is reserved for each \kmer to indicate which type of compression strategy is used.
If the difference does not fit into the smaller field, the entire extension information is exchanged.
An encoder on the sending side calculates the difference and decides whether to compress or not, while a decoder on the receiving side analyses the information.
This can also be overlapped with the communication.

The overlapping of computation and communication led to a $ 1.4\times$ speedup.
The compression strategy reduced the communication volume by $50\%$ and led to a further speedup of about $1.3\times$ when extension information is included.

\subsection{Task Abstraction Layer}
\label{sec:task_abstract}


To support flexible hybrid parallelism, we have chosen a parallel radix sort algorithm that can take advantage of multithreading.
However, both RADULS and PARADIS exhibit poor weak scaling performance once the number of threads exceeds 16 in our experiments.
To address this problem, we introduce task parallelism in \ours, which serves as an abstraction layer between processes and threads.
Distributed memory \kmer counters usually partition \kmers into $p$ batches, where $p$ is the number of processes, similar to a distributed hash table.
Conversely, in \ours, we partition the \kmers into $s$ batches, where $s$ is a parameter chosen at runtime.
Each batch is also referred to as a \emph{task}, and each task is treated as an independent processing unit, ensuring that \kmers with the same value can never belong to two different tasks.
Each task can be processed independently during the sorting and linear scan phases.

In the last stage, in which \kmers are sorted and counted, the available computing resources of a process are assigned to workers.
A worker can initiate several OpenMP threads, with each thread bound to a physical core.
Each worker is assigned some tasks and sorts them using the available resources.
Each worker processes the tasks independently, as shown on the right in Figure~\ref{fig:hysortk}.


By implementing hybrid task-based parallelism, we can utilize the numerous physical cores of modern CPUs.
By not starting too many processes on a single node or using fine-grained parallelization at the thread level, we can reduce scheduling overhead~\cite{bull1999measuring}.
In addition, instead of using all available threads to sort an array, which can lead to sublinear weak scaling, we divide the available cores into workers.
The number of threads per worker is constant.

The flexibility of task parallelism is also beneficial for the NUMA architecture of modern machines: threads in different NUMA nodes can be assigned to different tasks, eliminating implicit communication between NUMA nodes.
In practice, it is recommended to use at least one MPI process per NUMA domain, since the threads of a process share certain data structures, such as the MPI receive buffer.
The default number of threads per worker is set to $4$, which empirically provides great overall performance and flexibility.

Compared to the extreme cases where only one process or one process per physical core is used, the hybrid approach of \ours implemented using the task abstraction layer achieves a speedup of up to $5.7\times$ on a 128-core machine.

\subsection{Optimized Load Balance}
\label{sec:load_imbalance}

It is crucial to avoid load imbalances when implementing \kmer counting in distributed memory.
If using a bulk-synchronous parallel programming model such as MPI, any load imbalance on one process can cause unnecessary wait times for other processes, resulting in an overall slowdown.
It is therefore important to distribute the workload evenly across the processes.
The task abstraction of \ours provides greater scheduling flexibility thus eliminating low-to-medium load imbalances.
Despite this, load imbalance can still occur with certain input data.

Supermers are constructed and grouped, and the root process retrieves data about the size of each task before assigning it to a target process.
The supermers are then routed for processing based on the assignment.
The goal is to minimize the largest sum of task sizes for a single process, similar to solving the NP-complete Partition Problem~\cite{karp2010reducibility}.
In \ours, we use a greedy approach to obtain an approximately optimal assignment.
First, we define an upper threshold for the sum of task sizes that is close to the average size per process.
Then, we try to assign the tasks to the processes without exceeding the threshold.
If the assignment is not successful, we increase the threshold and try again.
The procedure is repeated until a successful assignment is made.

The hash function alone cannot guarantee that there is no severe load imbalance due to an inherent feature of DNA data, namely genomic repetition.
For example, the human genome contains numerous repeats of $(AATGG)_n$ \cite{jaishree1994human}.
This means that many of the \kmers extracted from these areas have the same value.
Consequently, regardless of the scoring function used, the \supermers created from these areas will go to the same target process, resulting in an unbalanced load in both the exchange and counting stages.
These frequent \kmers are often referred to as ``heavy hitters''.

In some studies, load imbalance in \kmer counting is addressed by customizing the ordering of \mmers, but this approach does not completely solve the problem.
Tools such as HipMer identify and treat the heavy hitters differently, which are first counted locally and then reduced globally~\cite{georganas2015hipmer}.
This reduces the imbalance during the local count, but the imbalance remains during the exchange phase as \kmers still need to be sent to their target process.
In addition, detecting and filtering out heavy hitters is resource-intensive.
Conversely, our task-based design provides a simple way to address this load imbalance.
\ours can collect statistical information about the size of each task and use this information as an indicator for the presence of potential heavy hitters.

If the task size reaches a certain threshold, which is typically the average task size multiplied by a constant, we conclude that the task contains heavy hitters and requires special handling.
Compared to identifying specific \kmers, identifying these tasks is effortless.
Once these heavy-hitter tasks are identified, we transform them locally, i.e. we no longer keep the \supermers, but the \kmers are extracted from the \supermers, sorted, and then counted locally.
Then, tuples of \texttt{(\kmer, count)}, which are referred to as \kmerlist representation, are sent to the target processes.
Once the target processes have received the \kmerlist, they perform another round of sorting and counting the tuples.
In the meantime, the other (non-heavy-hitter) tasks are processed normally.

Our strategy shown in Figure~\ref{fig:hysortk} solves the load imbalance in both the exchange and counting stages.
In the exchange stage, for heavy-hitter tasks, the \supermer compression strategy is no longer applicable, and an additional field is used to store the count. 
The communication volume of a heavy-hitter task is therefore generally larger than that of a normal task.
However, the \kmerlist representation eliminates the need to send identical \kmers multiple times from the same process.
The task abstraction further mitigates the load imbalance in the exchange stage.
Finally, \kmerlist is beneficial for the counting stage, as fewer elements need to be sorted.

\section{Experimental Results}\label{sec:results}

Table~\ref{tab:datasets} summarizes the input data used in our experiments. 
\ours takes FASTA files as input.
Unless otherwise noted, the experiments were performed on the Perlmutter supercomputer (CPU node) at National Energy Research Scientific Computing (NERSC). 
Each CPU node has 512 GB of total memory and two 64-core AMD EPYC 7763 (Milan) CPUs. 
The nodes are connected via HPE Slingshot NIC with a 3-hop Dragonfly topology. 
IO time is not included except for the last test as it is beyond the scope of this paper and may vary due to the machine's storage system, network status, and overall IO traffic.
The lower bound of the valid \kmer frequency is set to 2 and the upper bound to 50.
The extension information of \kmers is not included unless otherwise specified.

\begin{table}[t]
\centering  
\small  
\setlength\tabcolsep{7.5pt}  
\begin{tabular}{|l|r|}
\hline
Dataset name &  Size (GB) \\
\hline
\emph{A. baumannii} & 0.2 \\
\hline
\emph{C. elegans} & 4.5   \\
\hline
\emph{Citrus} & 17.0\\
\hline
\emph{H. sapiens} 10x & 31.0 \\
\hline
\emph{H. sapiens} (Short Read) & 36.0  \\
\hline
\emph{H. sapiens} 52x & 156.0\\
\hline
\end{tabular}
\vspace{.5em}
\caption{A summary of the data used in Section~\ref{sec:results}.
Data are publicly available at \url{https://portal.nersc.gov/project/m1982/bella/} and \url{https://www.ncbi.nlm.nih.gov/sra/ERX009609}.}
\vspace{-2em}
\label{tab:datasets}
\end{table}

\subsection{Optimization Strategies}\label{sec:optimization_strategies_perf}

\subsubsection{Task Abstraction Layer}

Using \emph{H. sapiens} 52x, we evaluated the optimization strategies proposed in Section~\ref{sec:task_abstract} and~\ref{sec:load_imbalance} to understand their impact on performance.
First, we used a baseline that combined \supermer and sorting-based counting and completed the task in $26.5$ seconds.
Then we included the task abstraction layer, which was designed to address the inefficiencies in processing low to medium load imbalances.
This approach split the workload across multiple tasks and assigned multiple workers per process, resulting in a reduced runtime of $23.4$ seconds, that is $1.13\times$ speedup, and demonstrating improved load imbalance handling.

A more significant improvement was achieved through the integration of our heavy hitter strategy.
This fully optimized approach, which uses a flexible task abstraction layer, significantly outperformed previous approaches by completing in just $15.4$ seconds, a $1.72\times$ speedup.
Using this approach, tasks that are most likely to contain heavy hitter \kmers are transformed from computation- and communication-intensive to communication-intensive only.
The task abstraction layer can then effectively address the load imbalance in the communication stage.

To further validate the impact of our task abstraction layer and the ability to implement fine-grained hybrid parallelism, we evaluate the impact of the \texttt{avg\_task\_per\_worker} parameter.
Our results indicate that a higher number generally achieves better performance.
On the \emph{H. sapiens} 52x dataset using 32 nodes, setting \texttt{tpw} (tasks per worker) to 3 resulted in performance that was $1.59\times$ faster than when \texttt{tpw} was set to 1 and $1.12\times$ faster than when \texttt{tpw} was set to 2.
This illustrates the advantage of assigning more tasks per worker and validates the use of our task abstraction layer.
They also show that this abstraction has the potential to improve computational efficiency in complex data processing scenarios.

\begin{table}[t]
\centering  
\small  
\setlength\tabcolsep{7.5pt}  
\begin{tabular}{|l|c|c|c|c|c|}
\hline
Processes per node &  4 & 8 & 16 & 32 & 64\\
\hline
\emph{C. elegans} (2 nodes) & \hspace{.5em}6.26 & \hspace{.5em}3.13 & 2.62 & 2.50 & 2.46 \\
\hline
\emph{H. sapiens} 10x (4 nodes) & 20.70 & 10.95 & 9.31 & 8.89 & 8.68 \\
\hline
\end{tabular}
\vspace{.5em}
\caption{\ours end-to-end runtime (s) varying the processes per node.}
\vspace{-2em}
\label{tab:tab1}
\end{table}

\begin{table}[t]
\centering  
\small  
\setlength\tabcolsep{4.25pt}  
\begin{tabular}{|l|c|c|c|c|c|}
\hline
Batch size &  10,000 & 20,000 & 40,000 & 80,000 & 160,000 \\
\hline
\emph{Citrus} (4 nodes) & 1.59 & 1.83 & 1.68 & 1.49 & 1.46 \\
\hline
\emph{H. sapiens} 52x (32 nodes) & 4.83 & 4.47 & 4.12 & 4.00 & 4.21 \\
\hline
\end{tabular}
\vspace{.5em}
\caption{\ours communication time (s) varying batch sizes.
}
\label{tab:tab2}
\vspace{-2em}
\end{table}

\begin{table}[t]
\centering  
\small  
\setlength\tabcolsep{7.5pt}  
\begin{tabular}{|l|c|c|c|c|c|}
\hline
$m$ &  7 & 13 & 17 & 21 & 27 \\
\hline
\emph{C. elegans} (1 node) & 10.41 & 4.26 & 4.48 & 4.91 & \hspace{.5em}6.55 \\
\hline
\emph{H. sapiens} 10x (4 nodes) & 27.73 & 8.09 & 8.51 & 9.53 & 14.71 \\
\hline
\end{tabular}
\vspace{.5em}
\caption{\ours end-to-end runtime (s) varying $m$ at $k=31$.
}
\label{tab:tab3}
\vspace{-2em}
\end{table}

\subsubsection{Hybrid OpenMP and MPI}\label{sec:hybridresult}

To further evaluate the performance of our hybrid approach, we tested \ours under different PPN (process per node) scenarios.
The number of nodes is fixed in each test group.
All available cores are used in each test, i.e. $OMP\_THREADS\cdot MPI\_PROCESSES = 256\cdot NODE\_NUM$.
Table~\ref{tab:tab1} shows that the best performance is achieved with at least 16 processes per node on two different input data.
The performance decreases rapidly if less than 16 processes per node are used.
The results are consistent with the theory.
Each Perlmutter CPU node has two sockets, 8 NUMA domains, and 16 CCX (Core Complex) sharing the L3 cache.
By allocating at least one process on each CCX domain, the implicit communication between the domains is eliminated. 



\subsubsection{Batch Size and Overlapping}


In Section~\ref{sec:comm_optimize} we described how communication takes place in rounds.
The sending size (or batch size) is limited for each round.
The preparation of the sending buffer and the parsing of the receiving buffer overlap with communication, along with other calculations.
A small message size can lead to a high communication overhead, while a large message size can lead to a low overlap.
To find the best size, we tested the performance of \ours at different batch sizes.
Table~\ref{tab:tab2} shows that with $80,000$ we minimize unnecessary overhead while maximizing the effective use of our computational resources, which increases performance.

\subsubsection{Length of $m$}
The length of \supermers and their distribution depend largely on $m$, which influences the runtime, as shown in Table~\ref{tab:tab3} for $k=31$.
A smaller $m$ gives a lower probability that adjacent \kmers fall into the same task.
However, there is a trade-off between the load imbalance, if $m$ is too small, and the total communication volume, if $m$ is too large.
In Table~\ref{tab:tab3}, both the communication and the sorting phase suffer from an imbalanced task size at $m=7$.
Empirically, we found that $m=k/2$ is optimal for a smaller $k$, while for a larger $k$ a constant of $m=23$ has an ideal performance.



\subsection{Scaling Performance}

\begin{figure}[t]  
  \centering  
  \includegraphics[width=\columnwidth]{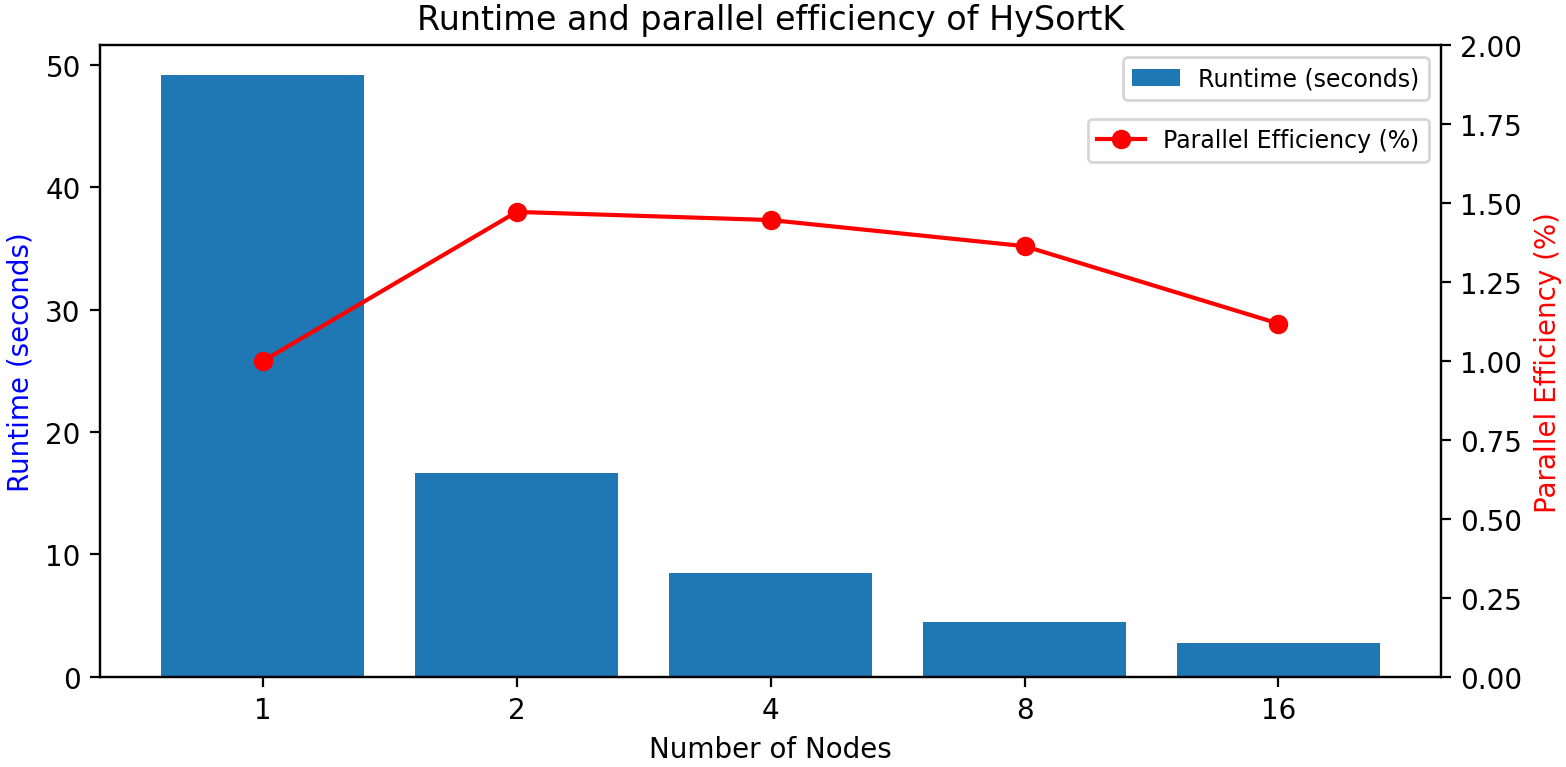}  
  \vspace{-2em}
  \caption{\ours strong scaling performance on the \emph{H. sapiens} 10x dataset using $k=31$.}  
  \Description{}
  \label{fig:humanstrongscaling}  
\end{figure}

\begin{figure}[t]  
  \centering  
  \includegraphics[width=\columnwidth]{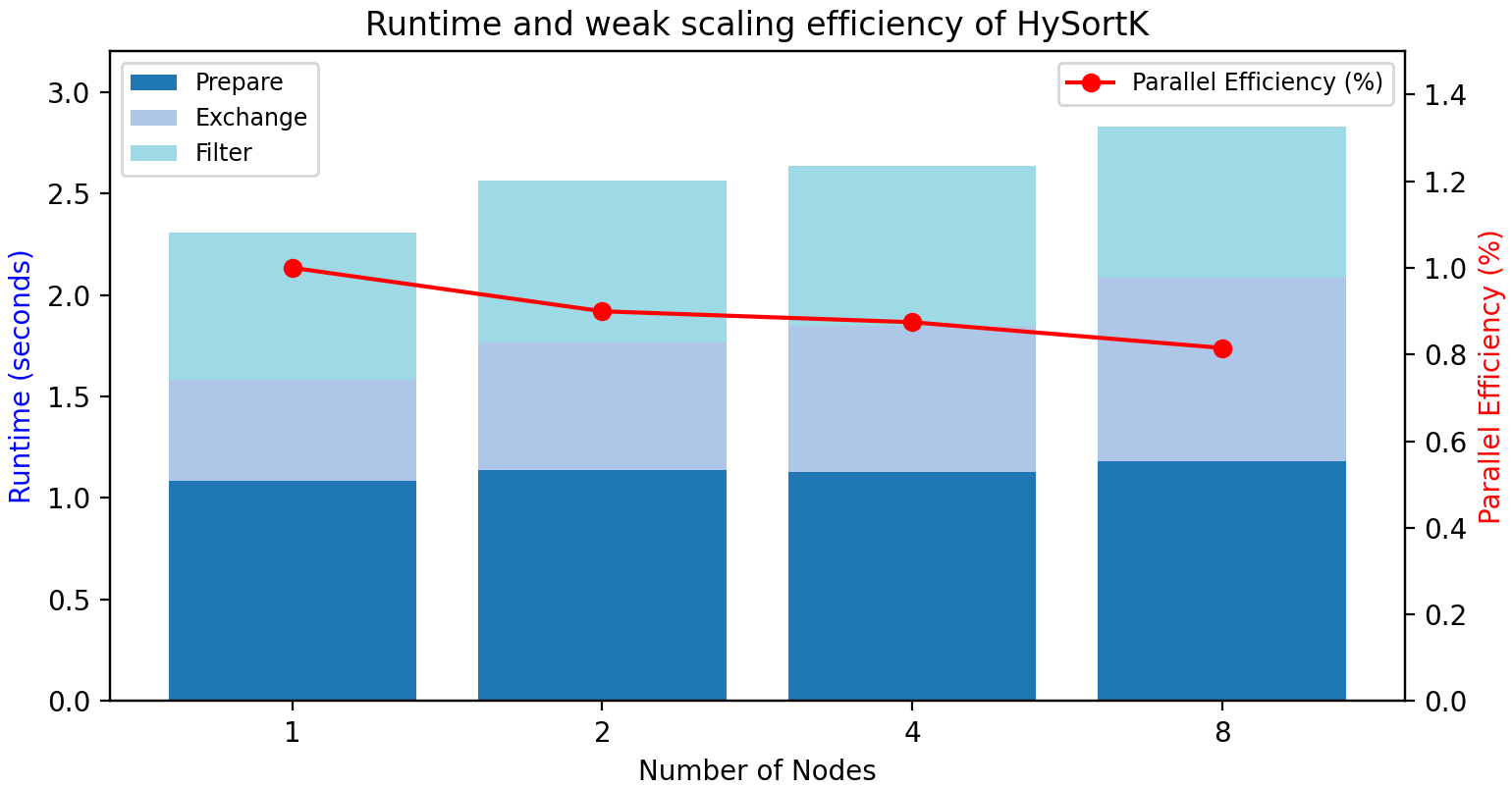}
  \vspace{-2em}
  \caption{\ours weak scaling performance using $k=31$.}  
  \Description{}
  \label{fig:humanweakscaling}  
\end{figure}

Figures~\ref{fig:humanstrongscaling} and~\ref{fig:humanweakscaling} show the overall performance of \ours using 16 processes per node (see Section~\ref{sec:hybridresult}), using all cores.
The circle marker indicates the scaling efficiency.
The scaling efficiency is high for smaller node counts and gradually decreases when the ratio of data size to core count is relatively small.
The decrease in efficiency is mainly due to the increase in communication.
On the \emph{H. sapiens} 10x dataset in Figure~\ref{fig:humanstrongscaling}, we achieve superlinear scaling efficiency.
On one node, we are forced to use the more memory-efficient but slower PARADIS sorting algorithm in \ours due to limited available memory.
Given more nodes, \ours automatically switches to the more efficient RADULS sorting algorithm.

\begin{figure}[t]  
  \centering  
  \includegraphics[width=\columnwidth]{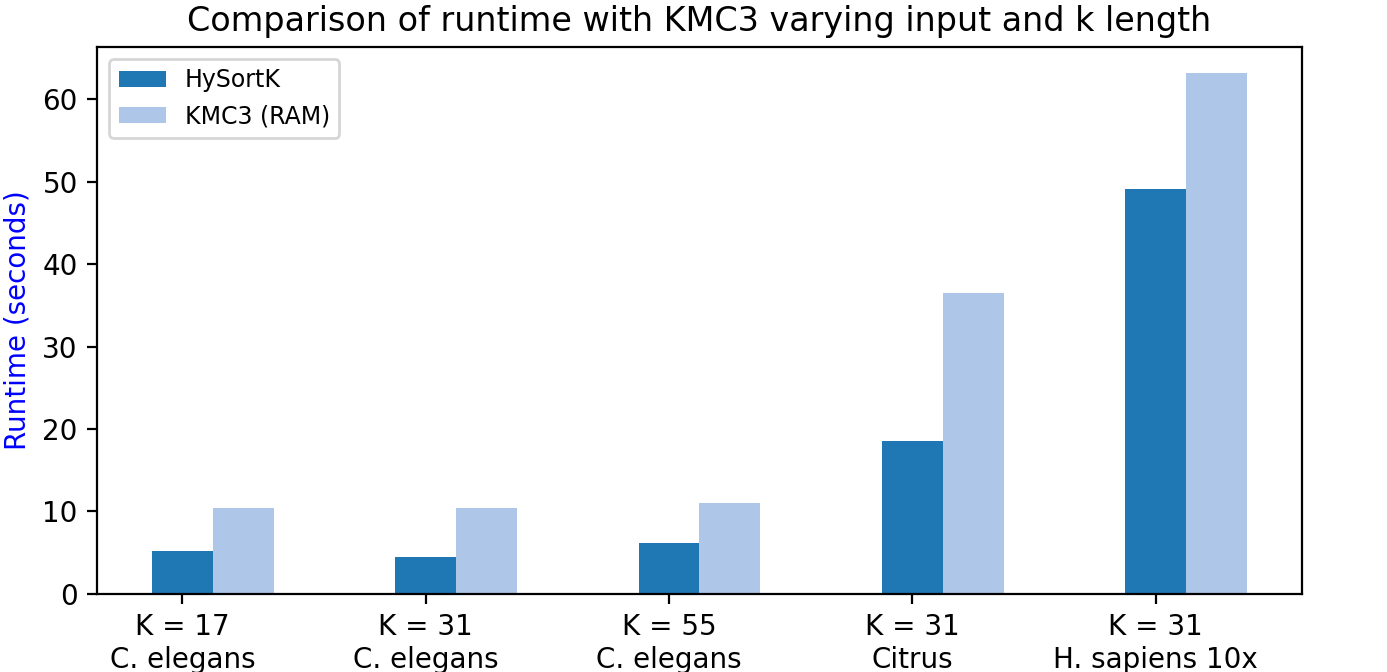}  
  \vspace{-2em}
  \caption{Comparison between \ours and KMC3~\cite{kokot2017kmc} on multiple input data and varying $k$ length.}  
  \Description{}
  \label{fig:kmc3comp}  
\end{figure}

Figure~\ref{fig:humanweakscaling} illustrates the weak scaling performance of \ours on the \emph{H. sapiens} (Short Read) dataset.
The preparation, exchange, and filter stages correspond to stage 1, stage 2, and stage 3 in the distributed \kmer counting paradigm (Section~\ref{sec:backgroundp}), respectively.
To perform this experiment, we divided the data set into different batches of 2 GB.
The number of batches increases proportionally with the number of nodes.
\ours achieved a weak scaling efficiency of about $80\%$ on 8 nodes.
Both the preparation and filtering stages exhibit perfect weak scaling, while the performance of the communication stage decreases as the number of nodes increases.
This is to be expected, as the processes run independently during the preparation and filtering phases.
In the communication phase, the average communication volume is constant, but the communication overhead increases as the number of nodes increases. 

\subsection{Shared-Memory Performance}\label{sec:sharedcompare}

To show the competitiveness of our approach, we compared \ours with KMC3, a popular shared-memory \kmer counter.
KMC3 does not support distributed memory parallelization.
It uses disk space to count \kmers when there is not enough RAM space to perform the computation on a single node.
To make the comparison fair, we use the \texttt{-r} option to force KMC3 to run in RAM-only mode, which does not allow the tool to use the hard disk.
The selected input data fits into RAM.
IO time is excluded for both KMC3 and our software.
The process per node is set to $16$, the batch size to $80,000$, and $m=\{11, 17, 23\}$ for $k=\{17, 31, 55\}$ respectively for \ours by default in Section~\ref{sec:sharedcompare} and~\ref{sec:distributedcompare}.

Figure~\ref{fig:kmc3comp} demonstrates we achieved competitive or better performance than KMC3 on various input data.
Both KMC3 and \ours use a sorting-based approach, so we conclude that the performance gain comes from our task-based parallelization scheme.
It provides flexibility, handles many available threads more efficiently, and copes better with NUMA or multi-socket scenarios.

\subsection{Distributed-Memory Performance}\label{sec:distributedcompare}

In this section, we compare \ours with kmerind and the \kmer analysis module of MetaHipMer2.
kmerind is a distributed memory CPU \kmer counter.
It uses optimized, cache-friendly hash tables to store the \kmers, while supporting the overlap of communication and computation.
For the experiments, we used the \texttt{ROBINHOOD, MURMUR64avx, CRC32C} variant of the improved kmerind \kmer counter as it shows the best performance~\cite{pan2018optimizing}. 
MetaHipMer2 is a \emph{de novo} meta-genome short-read assembler written in UPC++.
Its \kmer analysis module computes the \kmer count on GPUs and also uses the supermer strategy.
Despite our best efforts, we were unable to run DEDUKT on our machine.
SWAPCounter~\cite{ge2020counting} is excluded from the comparison due to its inaccurate output.

\begin{figure}[t]  
  \centering  
  \includegraphics[width=\columnwidth]{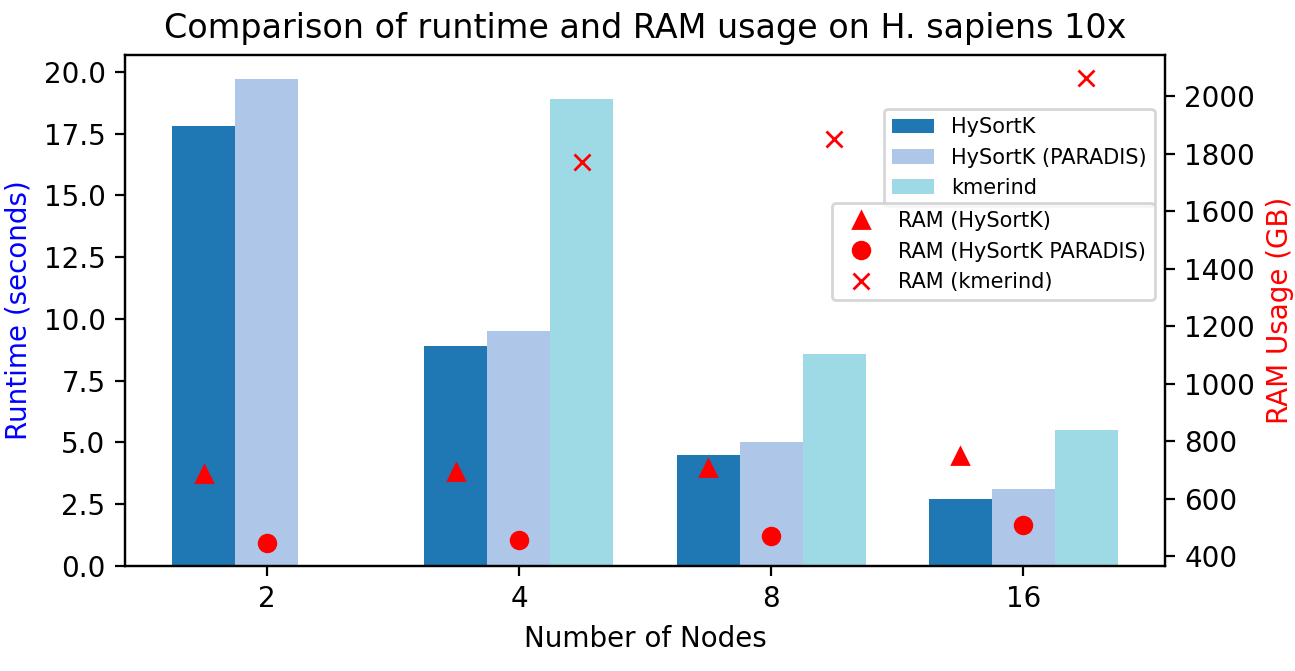}  
  \caption{Comparison of runtime and memory usage between \ours and kmerind~\cite{pan2018optimizing} on \emph{H. sapiens} 10x with $k=31$.}  
  \Description{}
  \label{fig:kmerindcomp-human10}  
\end{figure}

\begin{figure}[t]  
  \centering  
  \includegraphics[width=\columnwidth]{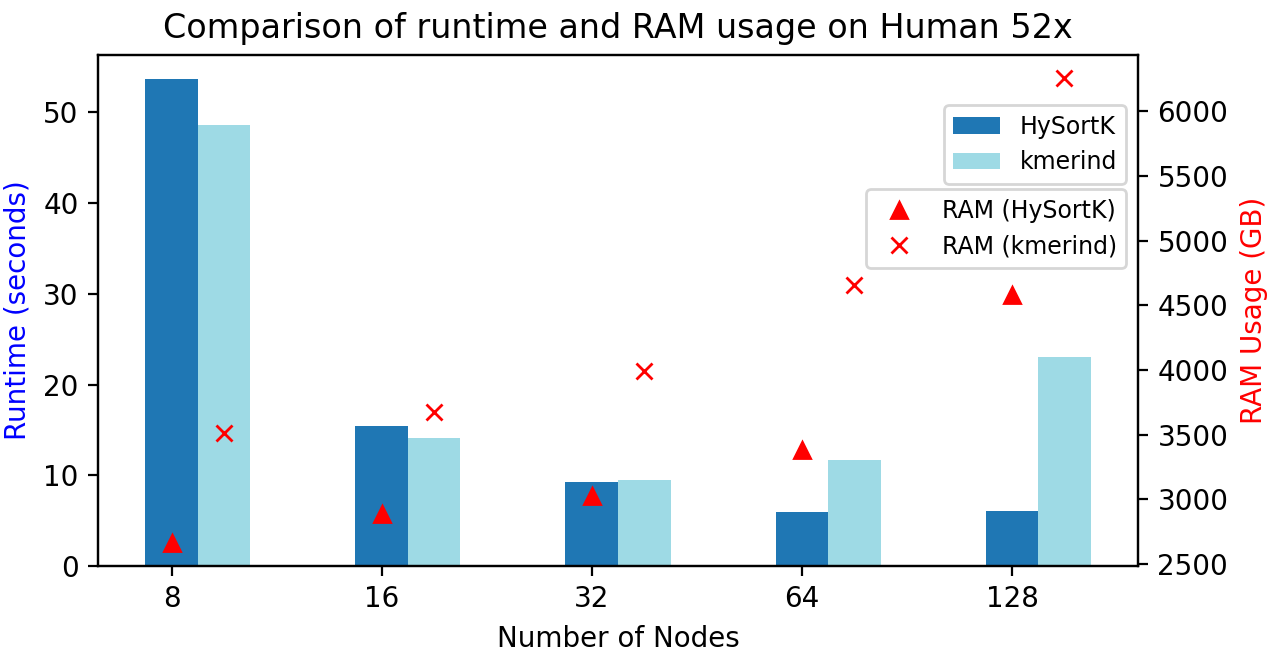} 
  \caption{Comparison of runtime and memory usage between \ours and kmerind~\cite{pan2018optimizing} with \emph{H. sapiens} 52x and $k=31$.}  
  \Description{}
  \label{fig:kmerindcomp-human52}  
\end{figure}

Figures ~\ref{fig:kmerindcomp-human10} and~\ref{fig:kmerindcomp-human52} report the runtime comparison between \ours and kmerind on two different datasets. 
The missing bar in Figure~\ref{fig:kmerindcomp-human10} indicates that kmerind ran out of memory under that setting.
Our \kmer counter is competitive or better than kmerind in every scenario.
Our RAM usage is 25\% to 70\% lower than kmerind's, as shown in Figure~\ref{fig:kmerindcomp-human10} on the right y-axis.
For the large \emph{H. sapiens} 52x dataset, \ours scales efficiently to 64 nodes thanks to the task layer and heavy hitter strategy, and does not suffer serious performance degradation as the number of nodes increases.
In contrast, kmerind can only scale to 32 nodes, and the runtime increases as the number of nodes increases.

\begin{figure}[t]  
  \centering  
  \includegraphics[width=\columnwidth]{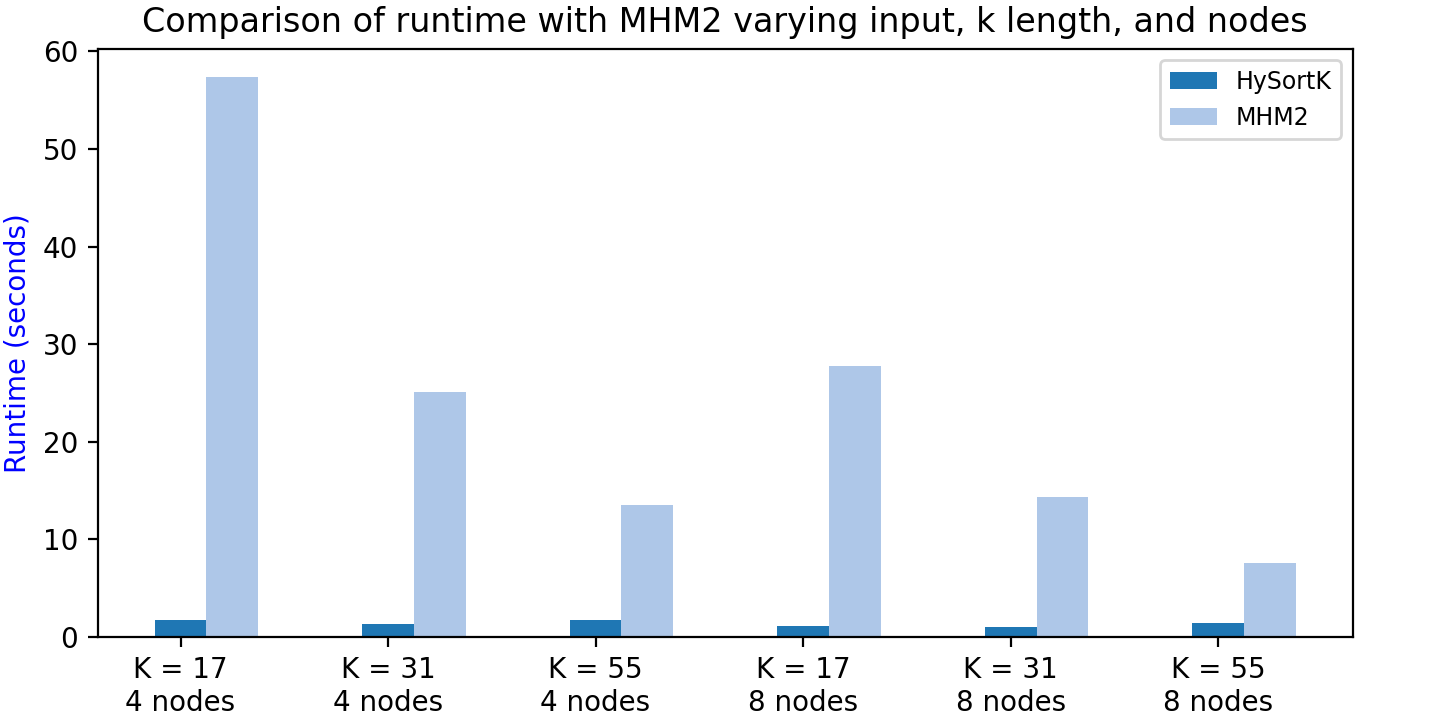}  
  \caption{Comparison between \ours and MHM2~\cite{awan2021accelerating} on the \emph{C. elegans} dataset varying $k$ length and number of nodes.}  
  \Description{}
  \label{fig:mhm2comp}  
\end{figure}

Figure~\ref{fig:mhm2comp} compares the runtime of \ours and MetaHipMer2 (MHM2) \kmer analysis on the \emph{C. elegans} dataset.
MHM2 is run on Perlmutter GPU nodes.
Each GPU node has 1 EPYC 7763 CPU, 4 NVIDIA A100 GPUs, and 4 HPE Slingshot11 NICs. 
\ours consistently outperformed MHM2 many times over.
Our hypothesis is that communication, including inter-CPU communication and CPU-GPU communication, is the bottleneck for MHM2.
The performance gap decreases as we increase the number of nodes.
As $k$ increases, the \supermers become longer, which reduces the total communication volume and leads to speedup.




\subsection{Integration into Real-World Pipeline}

To test our hypothesis that flexible hybrid parallelization can increase the efficiency of the entire pipeline, we integrated our \kmer counter into the distributed-memory ELBA \emph{de novo} long-read assembly pipeline.
ELBA's overlap detection, contig generation and transitive reduction modules already support hybrid OpenMP and MPI parallelization, while the \kmer counter stage does not use the hybrid parallelism, thus hindering the overall efficiency of the pipeline in shared memory scenarios.
ELBA is usually only executed with MPI processes and without OpenMP parallelization to account for the lack of multithreading parallelism in the \kmer counter.
In addition, ELBA requires not only the frequency of a specific \kmer, but also the origin of such a \kmer in the input sequences.
Therefore, we used the extension version of \ours.

\begin{figure}[t]  
  \centering  
  \includegraphics[width=\columnwidth]{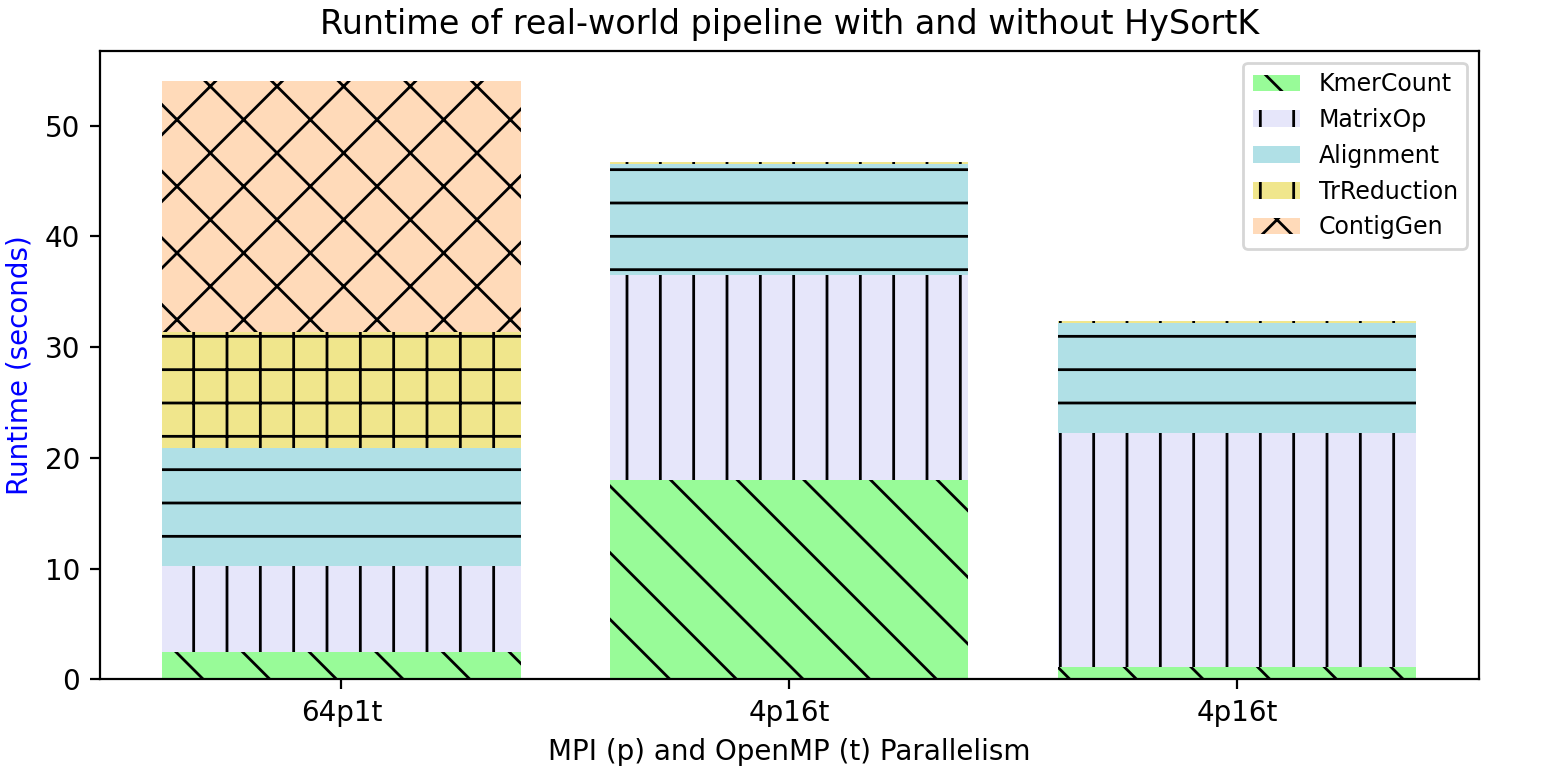} 
  \caption{ELBA with and without \ours at 64 processes and 1 thread (64p1t) or 4 processes and 16 threads (4p16t).}  %
  \Description{}
  \label{fig:elba}  
\end{figure}

Figure~\ref{fig:elba} illustrates the runtime breakdown with the \emph{A. baumannii} dataset on a single node using 64 cores.
Assigning one core per MPI process (no OpenMP parallelism) can lead to an increased runtime for transitive reduction and contig generation (left bar in Figure~\ref{fig:elba}).
On the other hand, using hybrid MPI and OpenMP parallelism makes \kmer counting much more time-consuming since the original \kmer counter in ELBA does not support OpenMP (middle bar in Figure~\ref{fig:elba}).
In contrast, we can leverage the hybrid parallelism when integrating \ours.
Our \kmer counter integrated into the ELBA pipeline significantly outperforms the original counter in a hybrid parallelization scenario (right bar in Figure~\ref{fig:elba}) and the original counter running exclusively under MPI.
Consequently, we achieve a speedup of 1.8$\times$ and 1.3$\times$ for the entire pipeline compared to the original pipeline running with 1 thread per process and 16 threads per process, respectively.

\section{Conclusions}\label{sec:conclusions}
Recent advances in high-throughput sequencing technologies have led to an exponential increase in the volume of genomic data.
In many genomic processing pipelines, \kmer counting is an important step and often the first, meaning that this step cannot escape the data size through a filtering process.
Therefore, highly parallel and memory-efficient \kmer counting on large-scale machines is crucial to accelerate genomic research.

In this work, we presented \ours, a highly efficient sorting-based \kmer counting algorithm implemented for distributed memory systems, which reduces random memory access and memory usage compared to conventional hash table-based approaches.
In addition, we introduced an improved supermer strategy to reduce the communication volume and the task abstraction layer to improve load balancing.
\ours consistently outperformed both shared memory and distributed memory software, even when using GPUs with a speedup of $2\times$ to $10\times$ and with up to $3\times$ less memory.
Future work includes implementing the supermer strategy with the \kmerlist task to further reduce the communication volume.

Finally, we showed how the task abstraction layer facilitates parallelism and improves performance; we believe that such abstraction has the potential to improve computational efficiency in complex data processing scenarios.

\begin{acks}
The work was developed during the first author's participation in an exchange program at Cornell University.
This project received support from the Center for Research on Programmable Plant Systems under National Science Foundation Grant No. DBI-2019674.
This research used resources of the National Energy Research Scientific Computing Center, a DOE Office of Science User Facility supported by the Office of Science of the U.S. Department of Energy under Contract No. DE-AC02-05CH11231 and award DDR-ERCAP0027296.
\end{acks}

\bibliographystyle{ACM-Reference-Format}
\bibliography{ref}

\end{document}